# Towards a Probabilistic Definition of Seizures

Ivan Osorio[1*], Alexey Lyubushin[2], Didier Sornette[3]

[1*] Corresponding Author, Department of Neurology, University of Kansas Medical Center, 390 Rainbow Boulevard, Kansas City, Kansas 66160
913 5884529 office
913 5884585 fax
iosorio@kumc.edu

[2] Institute of Physics of the Earth, Russian Academy of Sciences, 123995, Russia, Moscow, B.Gruzinskaya, 10

[3]ETH Zurich, Chair of Entrepreneurial Risks, D-MTEC, D-PHYS and D-ERDW, Kreuzplatz 5, CH-8032 Zurich, Switzerland

## Abstract

This writing: a) Draws attention to the intricacies inherent to the pursuit of a universal seizure definition even when powerful, well understood signal analysis methods are utilized to this end; b) Identifies this aim as a multi-objective optimization problem and discusses the advantages and disadvantages of adopting or rejecting a unitary seizure definition; c) Introduces a *Probabilistic Measure of Seizure Activity* to manage this thorny issue.

The challenges posed by the attempt to define seizures unitarily may be partly related to their fractal properties and understood through a simplistic analogy to the so-called “Richardson effect”. A revision of the time-honored conceptualization of seizures may be warranted to further advance epileptology.

The task of automated detection of epileptic seizures is intimately related to and dependent on the definition of what is a seizure, definition which to date is subjective and thus inconsistent within and among experts [1,2,3]. The lack of an objective and universal definition not only complicates the task of validation and comparison of detection algorithms, but possibly more importantly, the characterization of the spatio-temporal behavior of seizures and of other dynamical features required to formulate a comprehensive epilepsy theory.

The current state of automated seizure detection is, by extension, a faithful reflection of the power and limitations of visual analysis, upon which it rests. The subjectivity intrinsic to expert visual analysis of seizures and its incompleteness (it cannot quantify or estimate certain signal features, such as power spectrum) confound the objectivity and reproducibility of results of signal processing tools used for their automated detection. What is more, several of the factors, that enter into the determination of whether or not certain grapho-elements should be classified as a seizure, are non-explicit ("gestalt-based") and thus difficult to articulate, formalize and program into algorithms. Most, if not all, existing seizure detection algorithms are structured to operate as expert electroencephalographers. Thus, seizure detection algorithms that apply expert-based rules are at once useful and deficient; useful as they are based on a certain fund of irreplaceable clinical knowledge and deficient as human analysis biases propagate into their architecture. These cognitive biases which pervade human decision processes and which have been the subject of formal inquiry [4-6] are rooted in common practice behaviors such as: a) The tendency to rely too heavily on one feature when making decisions (e.g., if onset is not sudden, it is unlikely to be a seizure because these are paroxysmal events); b) To declare objects as equal if they have the same external properties (e.g., this is a seizure because it is just as rhythmical as those we score as seizures) or c) Classify phenomena by relying on the ease with which associations come to mind (e.g., this pattern looks just like the seizures we reviewed yesterday).

The seizure detection algorithms' discrepant results (Osorio et al, this issue) makes attainment of a unitary or universal seizure definition ostensibly difficult; the notion that expert cognitive biases are the main if not only obstacle on the path to "objectivity" is rendered tenuous by these results. These divergences in objective and reproducible results may be attributable in part, but not solely, to the distinctiveness in the architecture and parameters of each algorithm. The fractal or multi-fractal structures of seizures [7,8] accounts at least in part for the differences in results and draws attention to the so-called "Richardson effect". Richardson [9] demonstrated that the length of borders between countries (a natural fractal) is a function of the size of the measurement tool, increasing without limit as the tool's size is reduced. Mandelbrot, in his seminal contribution "How long is the coast of Britain" [10] stressed the complexities inherent to the Richardson's effect, due to the dependency of particular measurements on the scale of the tool used to perform them. Although defining seizures as a function of a detection tool would be acceptable, this approach may be impracticable when comparisons between, for example, clinical trials or algorithms are warranted. Another strategy to bring unification of definitions is to universally adopt the use of one method, but this would be to the detriment of knowledge mining from seizure-time series and by extension to clinical epileptology.

A Probabilistic Measure of Seizure Activity (*PMSA*) is proposed as one possible strategy for characterization of the multi-fractal, non-stationary structure of seizures, in an attempt to eschew the more substantive limitations intrinsic to other alternatives.

The *PMSA* relies in this application on "indicator functions" (IFs) denoted $\chi_{\mathrm{algo}}$ for each algorithm 'algo' and also on an *Average Indicator Function (AIF)*:

$$AIF(t) = (\chi_{Val}(t) + \chi_{r^2}(t) + \chi_{STA/LTA}(t) + \chi_{WTMM}(t)) / 4$$

The subscripts Val, $r^2$, STA/LTA and WTMM refer to four different algorithms described briefly below and more extensively in (Osorio et al, 2011, this issue) [Ivan Osorio, Alexey Lyubushin and Didier Sornette, Automated Seizure Detection: Unrecognized Challenges, Unexpected Insights, Epilepsy & Behavior (2011)]. An algorithm's IF equals 1 for time intervals (0.5 sec in this application) "populated" by ictal activity and 0 by inter-ictal activity. The IF's are used to generate four stepwise time functions, one for each of: a) A $2^{nd}$order auto-regressive model ($r^2$); b) The Wavelet Transform Maximum Modulus (*WTMM*); c) The ratio of short-to-long term averages (*STA/LTA*) and d) The Validated algorithm (*Val*). With these IFs, the *AIF* is computed (its values may range between [0-1] with intermediate[1] values of 0.25, 0.5 and 0.75 in this application). These values [0-1] are estimates of the probability of seizure occurrence at any given time.

The dependencies of *AIF* values on the detection algorithm applied to the ECoG are illustrated in figure 1a-d and reflect the probability that grapho-elements are ictal in nature; the higher the *AIF* value, the greater the probability that the detection is a seizure. *AIF* values of 1 (the activity is detected by all algorithms as a seizure) or 0 (none of the algorithms classifies the grapho-elements as a seizure) pose no ambiguity, but as shown in this study, are likely to be less prevalent than intermediate values [$0 < AIF < 1$]. By way of example, cortical activity may be classified as a seizure if the *AIF* value is 0.75, having been detected by the majority (¾) of methods. In the study published in this issue (Osorio et al.), the four different methods ($r^2$, WTMM, STA/LTA, and Val) were investigated, but this number may vary according to the task at hand; for warning for the purpose of allowing operation of a motor vehicle, application of a larger number of detection algorithms to cortical signals and an *AIF* value of 1 would be desirable while, for automated delivery of an innocuous, power inexpensive therapy, less algorithms and much lower *AIF* values would be tolerable.

The cross-correlation between each pair of algorithm's IF and their average function (*AIF*) were calculated; since each of these is a step function (see figure 1), the Haar wavelet transform was applied to them to facilitate visualization of their value (*y*-axis) as a function of this wavelet's logarithmic time scale (*x*-axis (Figure 2). The correlations (indicative of the concordance level) between each IF pair and between each method's IF and the *AIF,* increases monotonically, reaching a maximum between 20-30s, after which they decrease also monotonically (except for *AIF* vs. $r^2$ ): The WTMM and $r^2$ methods have the highest correlations with *AIF* for time scales exceeding 100 sec. Since estimating the probability measure of seizure activity based on the *AIF* requires the output of at least two detection algorithms, a simpler approach is to apply only one, a Wavelet Transform Maximum Modulus-Stepwise Approximation (WTMM-*SAp*).

Let $U(\xi_j)$ be a logarithm of the standard deviation of differentiated ECoG computed within "small" adjacent time windows of length $L$ and $\xi_j$ the time moments corresponding to right-hand ends of these windows. Thus, $\xi_j$ values are given within the step $L \cdot \delta t$, where $\delta t$ is an ECoG time interval.

---

[1] Intermediate *AIF* values are functions of the number of algorithms applied to the signal. Since in this study 4 methods were used and the range of the indicator function is [0-1], the intermediated values are [0.25, 05, 0.75].

Let $S_U(\xi \mid a_*^{(j)})$ be a WTMM-*SAp* computed for the dyadic sequence of $m$ dimensionless scale thresholds:

$$a_*^{(j)} = a_*^{(0)} \cdot 2^{(j-1)}, \quad j = 1, ..., m \tag{26}$$

and $S_U^{(a)}(\xi)$ be their mean value:

$$S_U^{(a)}(\xi) = \sum_{j=1}^{m} S_U(\xi \mid a_*^{(j)}) / m \tag{27}$$

The averaged WTMM-*SAp* $S_U^{(a)}(\xi)$ may reveal abrupt changes of $U(\xi_j)$ for different scales (the use of a dyadic sequence (26) suppresses "outliers"). The background is estimated by a simple average within a moving time window of the radius of $n$ discrete values of $\xi$:

$$\bar{S}_U^{(a)}(\xi_j) = \sum_{k=-n}^{n} S_U^{(a)}(\xi_{j+k}) / (2n+1) \tag{28}$$

Seizures correspond to positive peaks of $S_U^{(a)}(\xi)$ above background $\bar{S}_U^{(a)}(\xi)$. Thus, the values:

$$\Delta S_U^{(a)}(\xi) = \max\{\, 0, S_U^{(a)}(\xi) - \bar{S}_U^{(a)}(\xi)\} \geq 0 \tag{29}$$

are regarded as a Measure of Seizure Activity (MSA). In order to make this measure probabilistic (PMSA), consider an empirical probability distribution function:

$$F_{\Delta S_U^{(a)}}(X) = \Pr\{\, \Delta S_U^{(a)}(\xi) < X \,\} \tag{30}$$

and let $Q_{\Delta S_U^{(a)}}(\gamma)$ be the $\gamma$-quantile of the function (30), i.e. the root of the equation:

$$F_{\Delta S_U^{(a)}}(Q) = 1 - \gamma, \quad 0 < \gamma < 1 \tag{31}$$

The *PMSA* is defined by the formula:

$$\mu(\xi) = \min\{\Delta S_U^{(a)}(\xi), Q_{\Delta S_U^{(a)}}(\gamma)\} / Q_{\Delta S_U^{(a)}}(\gamma), \quad 0 \leq \mu(\xi) \leq 1 \tag{32}$$

It should be underlined that the *PMSA* (32) is defined within sequences of "small" time intervals of length $L \cdot \delta t$ and $\xi = \xi_j$ are discrete time values, corresponding to right-hand ends of these time windows.

The method of constructing a *PMSA* based on the WTMM-*SAp* utilizes the following parameters whose values are shown in parentheses:

1) The number $L$ of adjacent samples for computing the logarithm of the standard deviations $U(\xi_j)$ for differentiated ECoG increments ($L = 240$).

2) The values of $a_*^{(0)}, m$ for setting the dyadic sequence of WTMM scale thresholds in the formula (26) ($a_*^{(0)} = 5, m = 6$, e.g., the following scale thresholds were used: 5, 10, 20, 40, 80 and 160).

3) The number $n$ of $\xi_j$ values for the radius of the moving averaging in formula (28) ($n = 200$, e.g., for $L = 240$ and $1/\delta t = 239.75$ Hz, the averaging length within formula (28) equals 401 sec).

4) The probability level $\gamma$ for calculating a quantile in formula (31) ($\gamma = 0.01$).

The results of the estimations of *PMSA* using WTMM-*SAp* (Figure 3) differ in one aspect (lower number of events with probability 1) from those obtained with the *PMSA-AIF*, given the dissimilarities between these two approaches, but are alike in uncovering the dependencies of *PMSA* on seizure duration: in general, the shorter the duration of a detection, the larger the discordance between detection methods, a "trait" that interestingly, is also shared by expert epileptologists (Osorio et al, 2002). Inter-algorithmic concordance as evidenced by the cross-correlation values between *PMSA*-WTMM-*SAp* and *PMSA-AIF* (Figure 4) grow quasi-linearly (albeit non-monotonically) with the temporal length of seizures, reaching a maximum value (0.73) at 250 s. Worthy of comment is the decay in cross-correlation values for seizure exceeding a certain length for both *PMSA-AIF* and *PMSA*-WTMM-*SAp*

The crafting of, or "convergence" towards, a unitary seizure definition would be epistemologically expensive and may thwart/delay deeper understanding of the dynamics of ictiogenesis and of the spatio-temporal behavior of seizures at relevant time-scales. In the absence of a universal definition, substantive gains are feasible through steps entailing, for example, the application of advanced signals analyses tools to ECoG, to hasten the identification of properties/features that would lead to the probabilistic discrimination of seizures from non-seizures with worthwhile sensitivity and specificity for the task at hand. Tools such as those available through cluster analysis of multidimensional vectors of relevant features would aid in the pursuit of automated seizure detection and quantification. To even have a modicum of success, this approach should not ignore the non-stationarity of seizures and strike some sort of balance between supervised (human) and unsupervised machine-learning) approaches. The resulting multidimensional parameter space, expected to be broad and intricate, may also foster discovery of hypothesized (e.g. pre-ictal) brain sub-states.

Seizure detection belongs to a class of optimization problems known as "multi-objective" [11] due to the competing nature between objectives; improvements in specificity of detection invariably degrade sensitivity and vice-versa. Attempts to achieve a universal seizure definition are likely to be fraught with similar competing objectives, but imaginative application of tools from the field of multi-objective optimization, among others, are likely to make this objective more tractable.

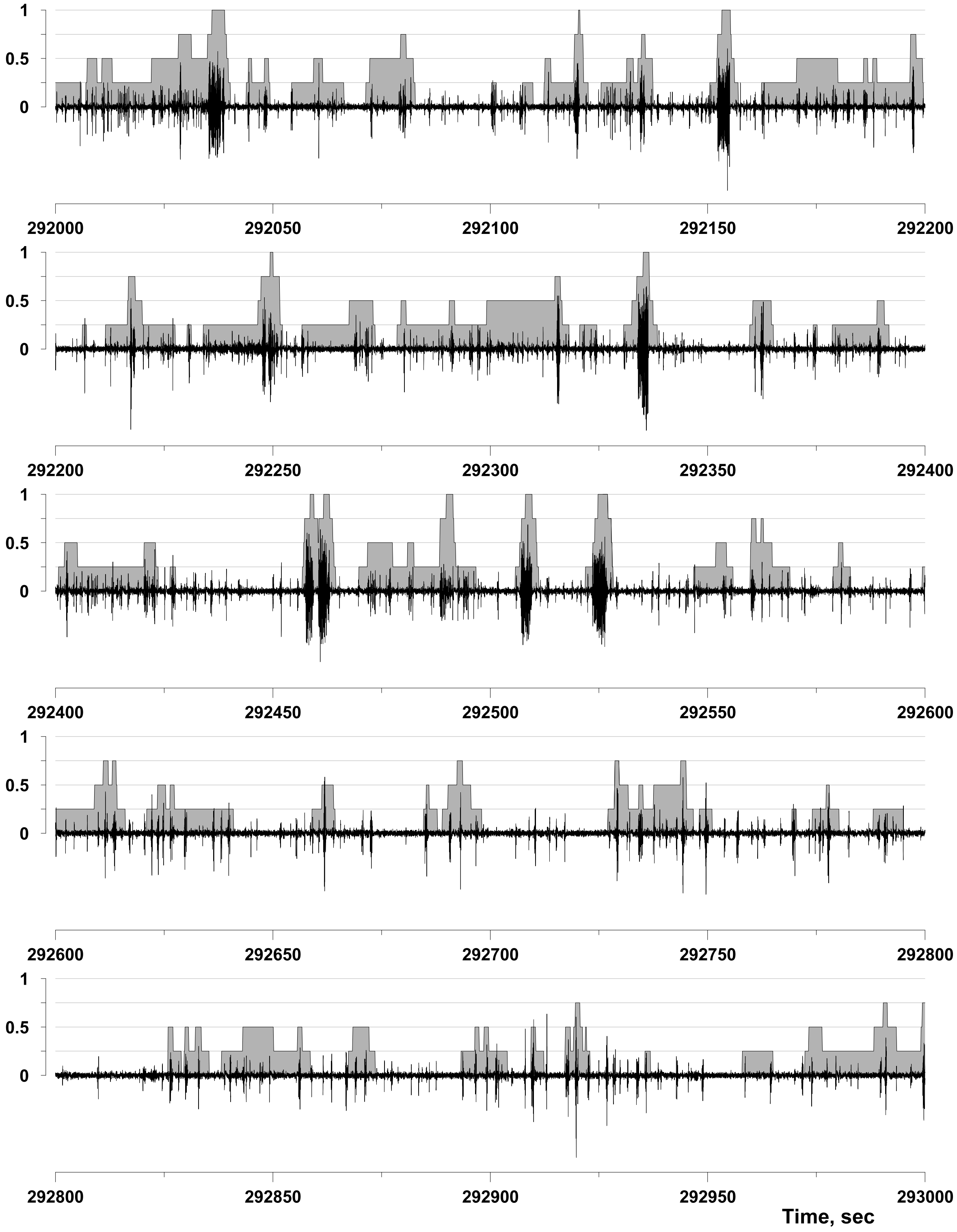

**Figure 1a**

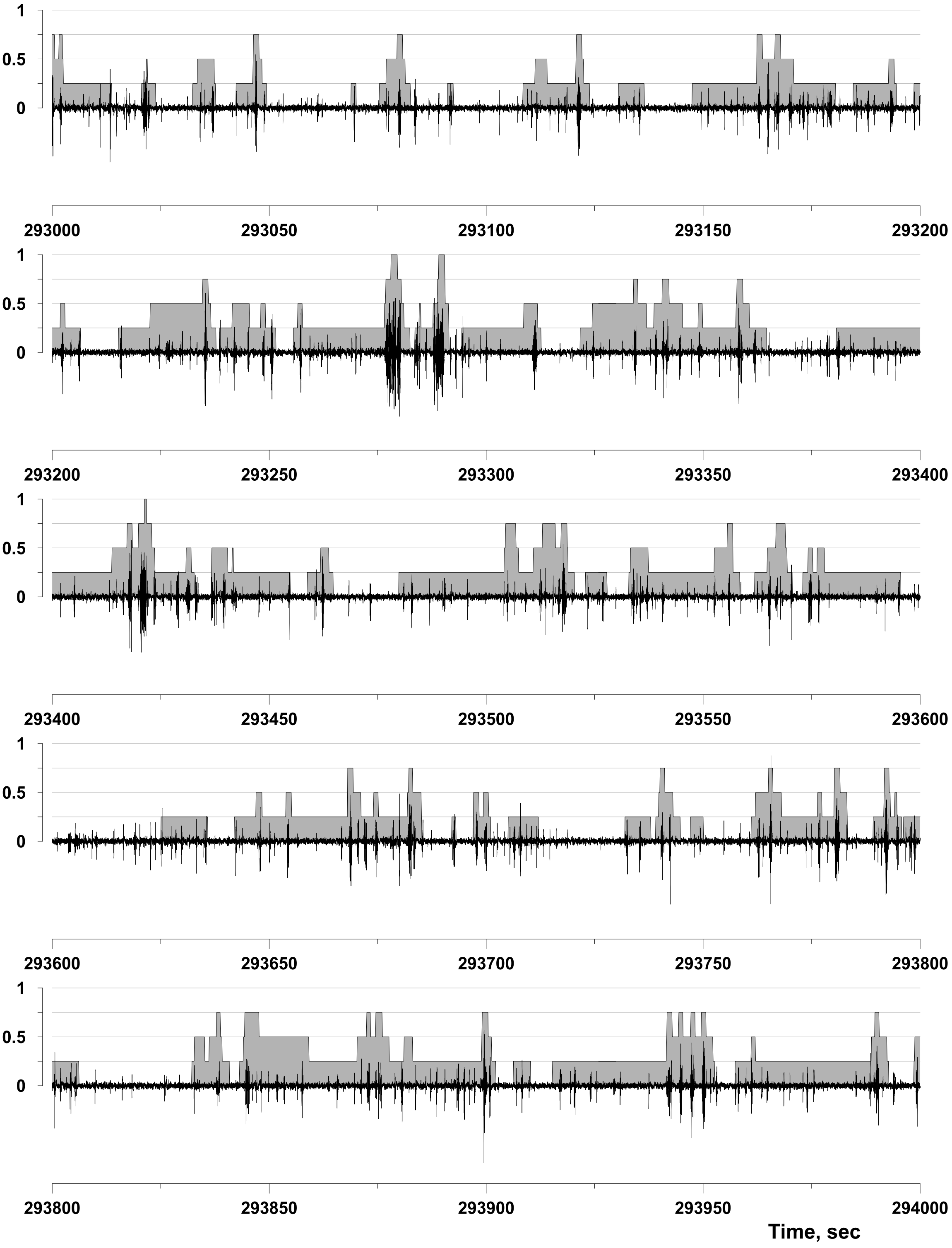

**Figure 1b**

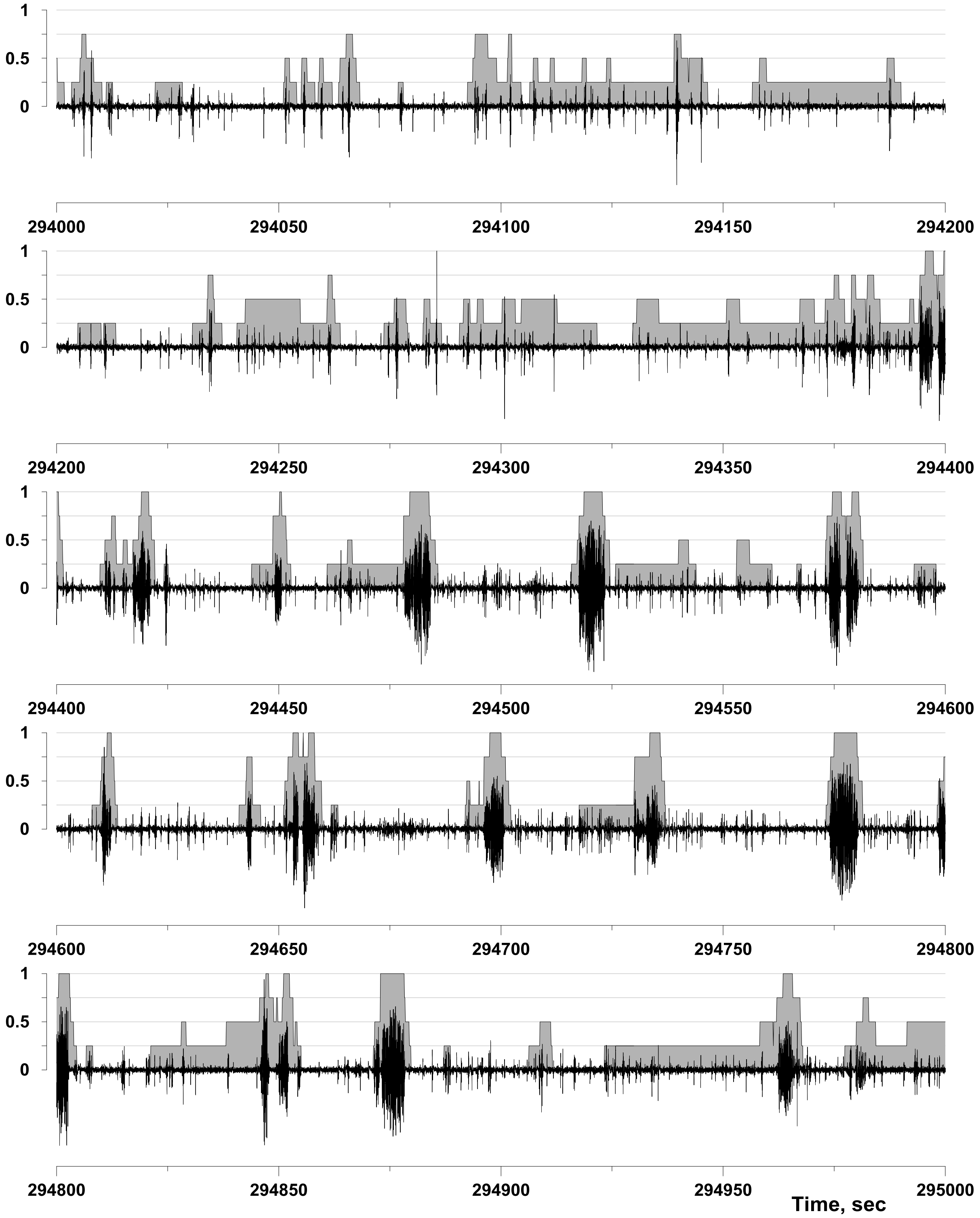

**Figure 1c**

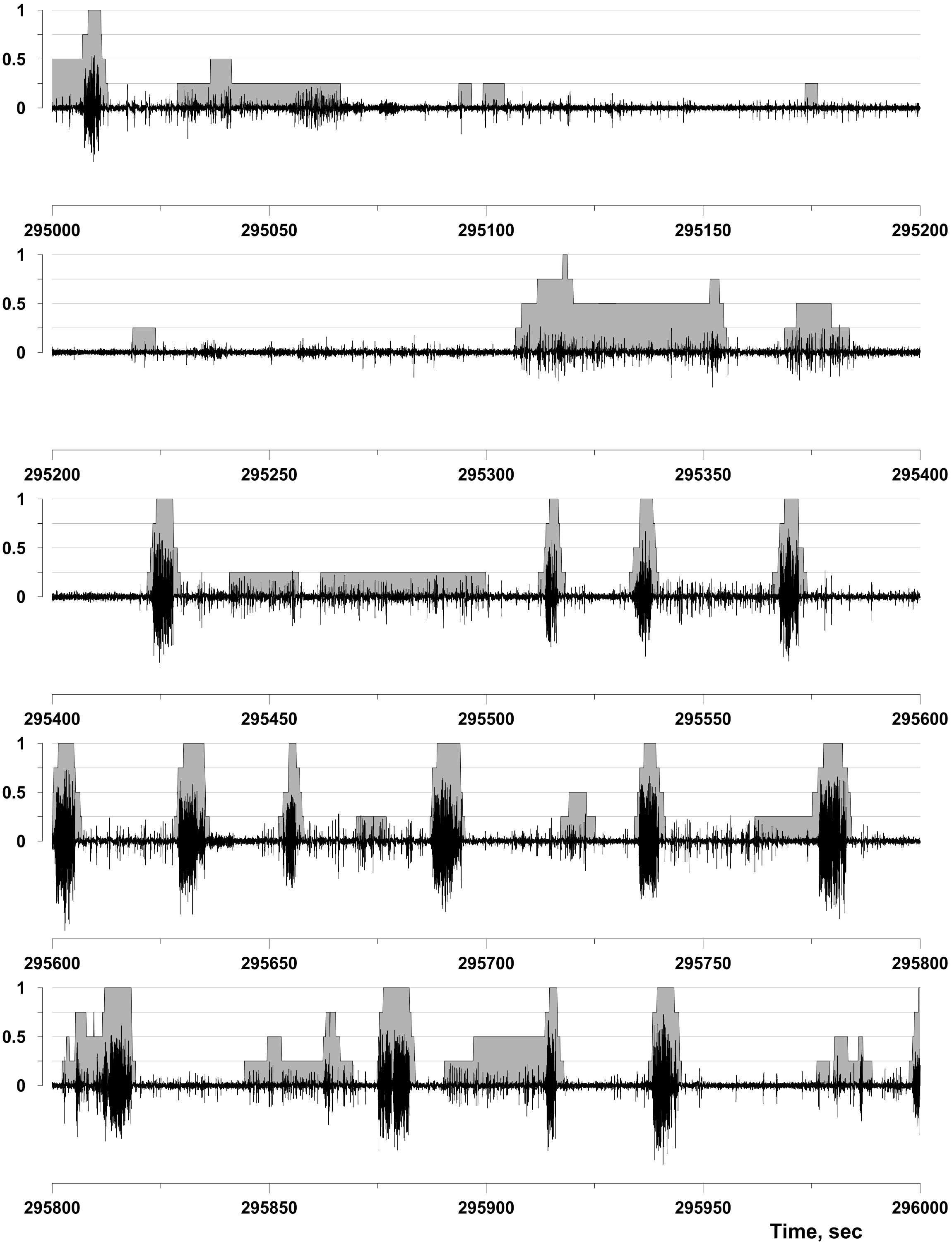

**Figure 1d**

**Figure 1a-d**. Average Indicator Function value (*AIF*; grey step-wise functions) of the probability that cortical activity (black oscillations) is a seizure over a certain time interval. The *AIF* value (0-1) of this function is calculated based on the output of each of the four detection algorithms used. Notice that the larger amplitude, longer oscillations are the only ones to have an *AIF* value of 1, indicative of "consensus" among all detection algorithms (*x*-axis: time; *y*-axis: *AIF* values)

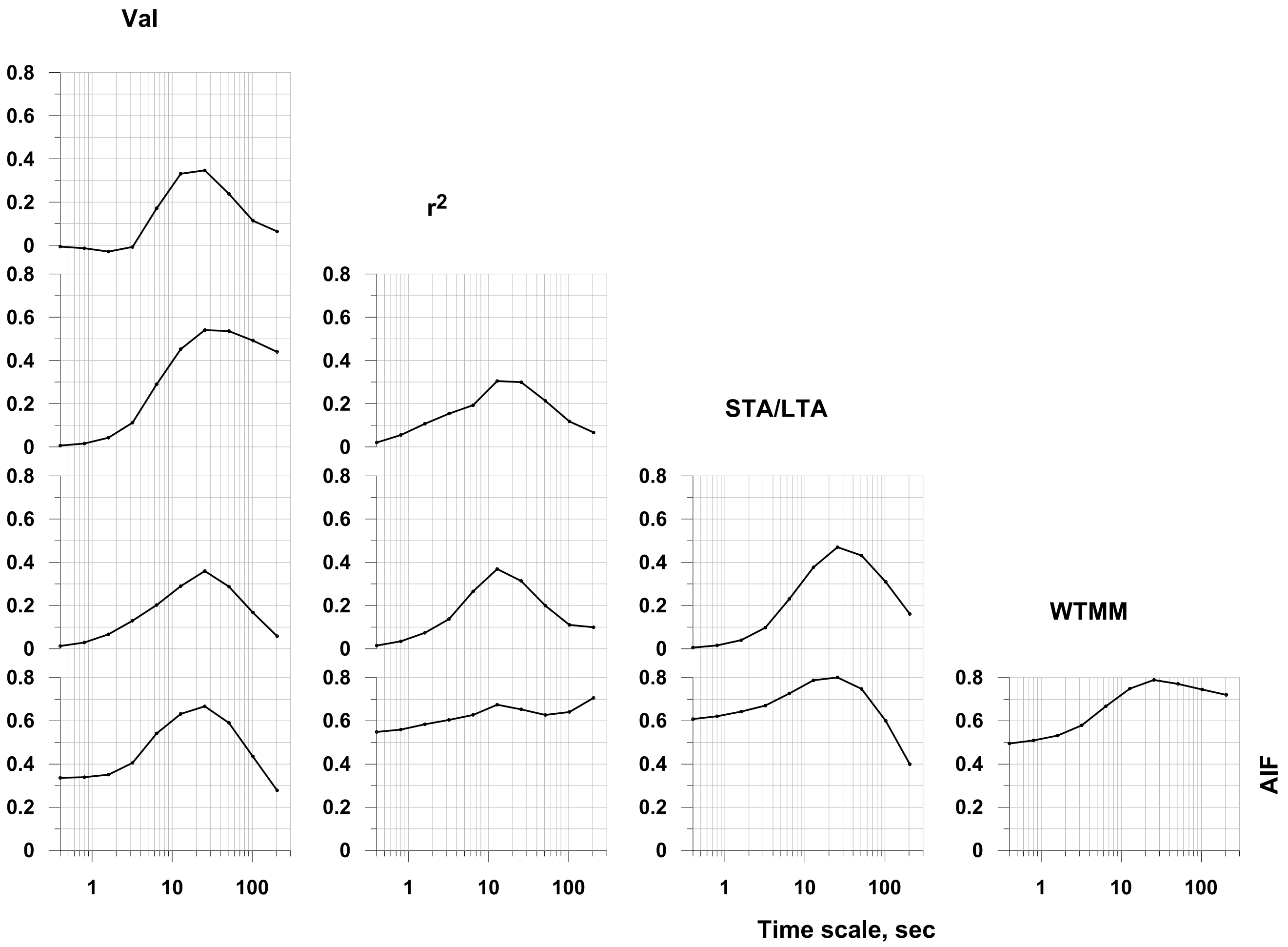

**Figure 2**. Plots of time scale-dependent correlations between Haar wavelet coefficients of the indicator functions (IFs) between pairs of detection methods and between each method and the averaged indicator function (*AIF*). Notice that $r^2$, *STA/LTA* and *WTMM* act as labels for both columns (label on top) and rows (label to the right of each row), whereas *Val* designates only the column below it and *AIF* the row to its left. This graph may be viewed as the lower half of a square matrix; this triangle's vertices are: the top left-most plot depicts the correlation between *Val* and $r^2$, the bottom left-most plot the correlation between *Val* and *AIF* and the bottom right-most graph, that between *WTMM* and *AIF;* all other correlations lie within these vertices (*y*-axes: Correlation values; *x*-axes: Logarithmic time scale).

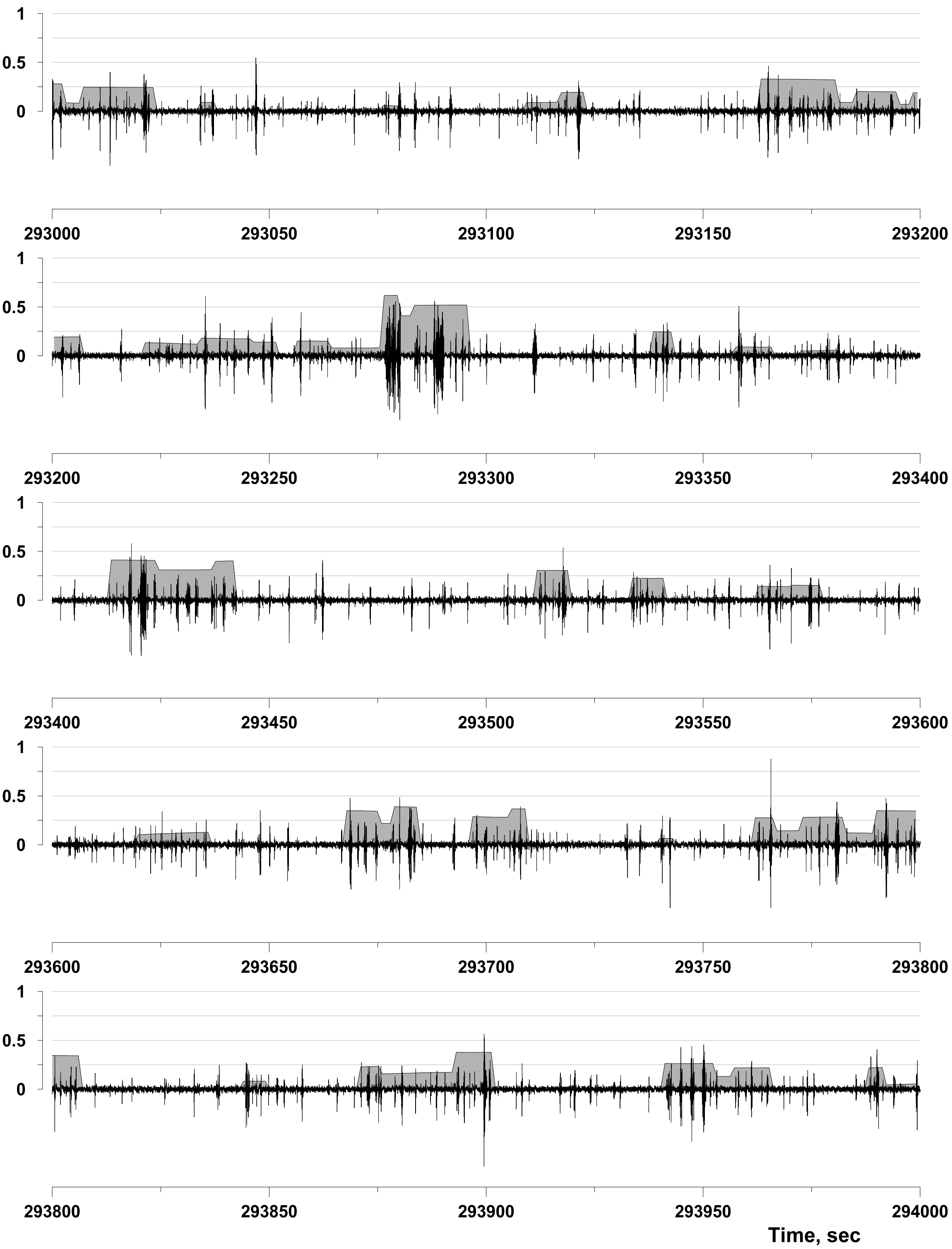

Figure 3a

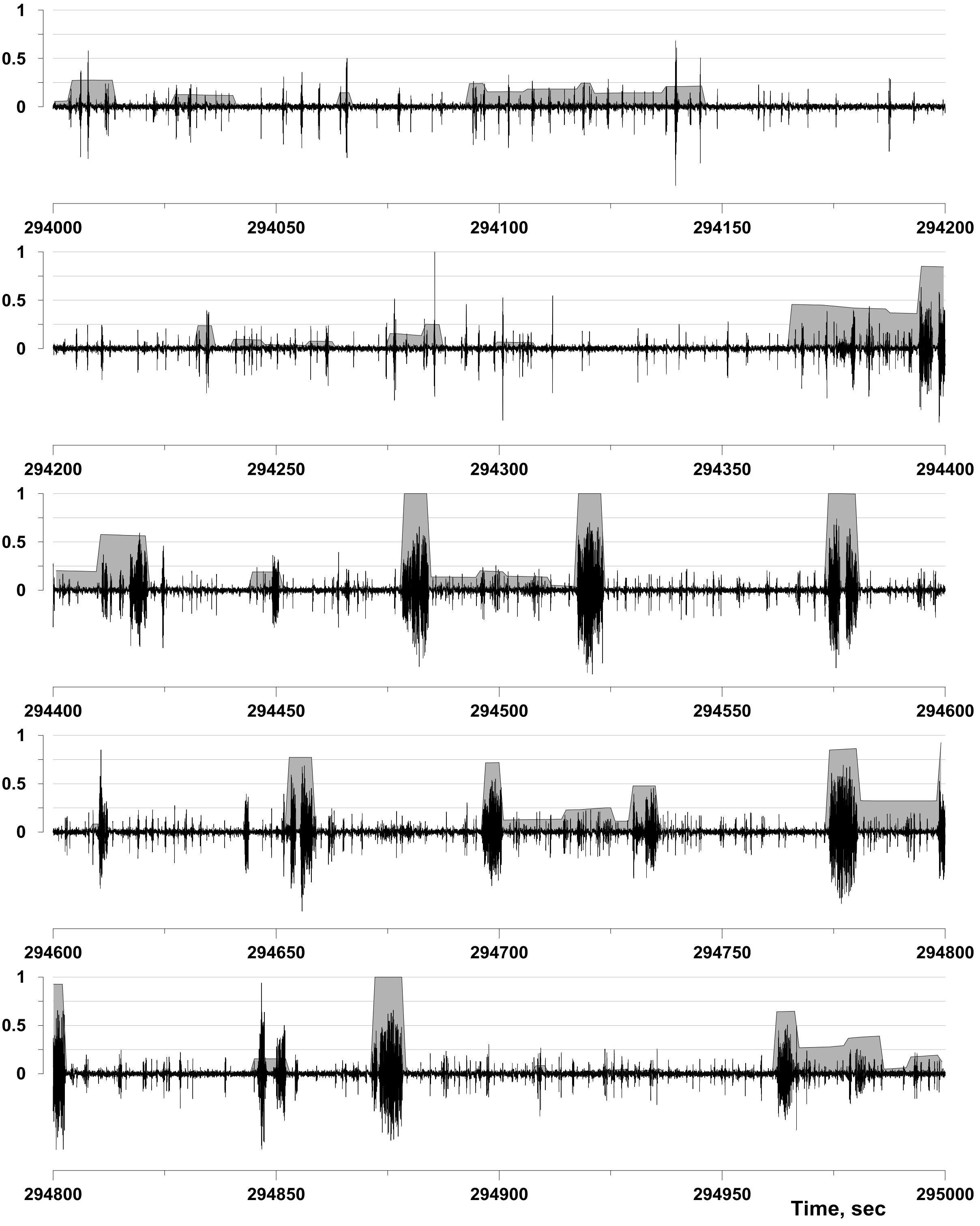

**Figure3b**

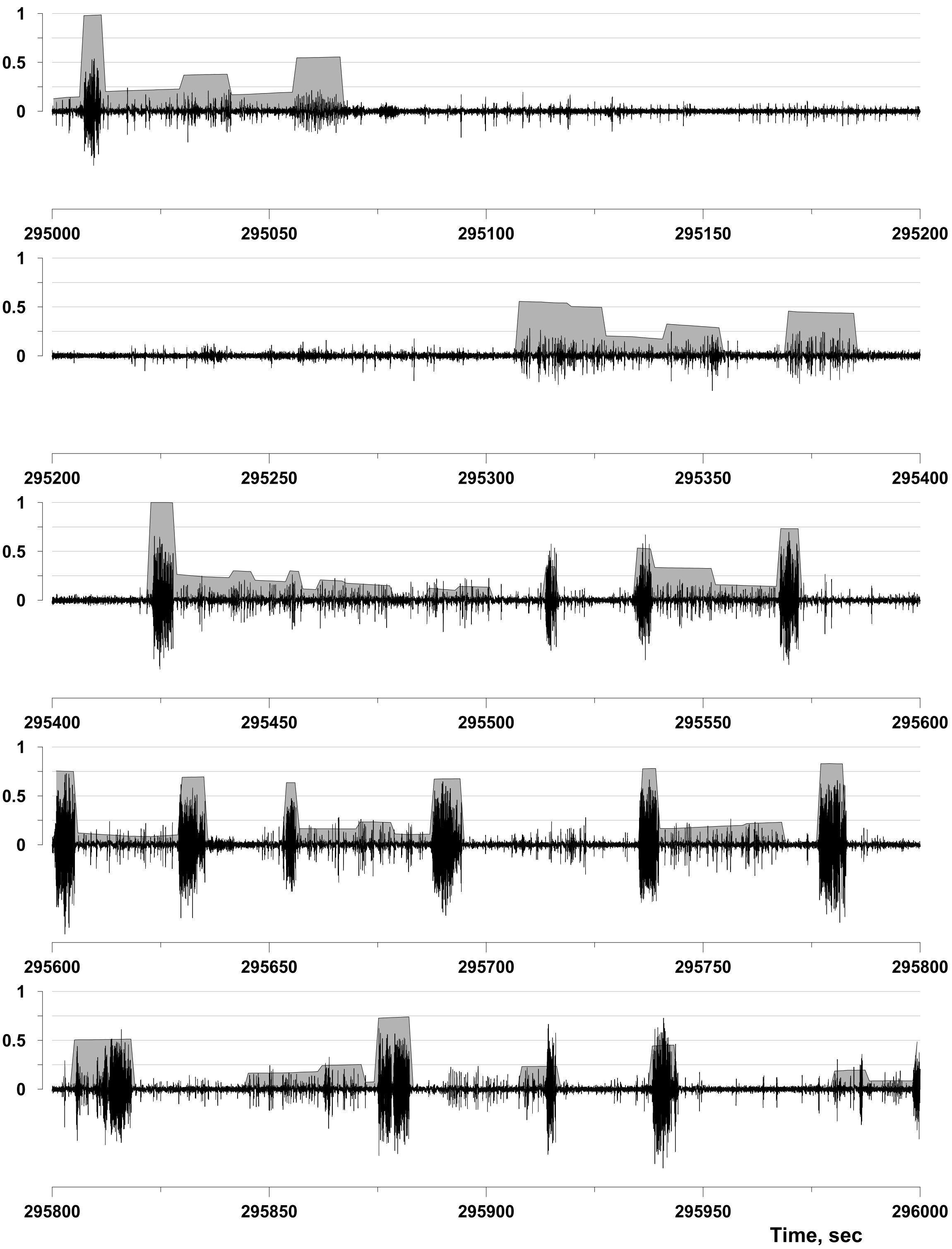

**Figure 3c**

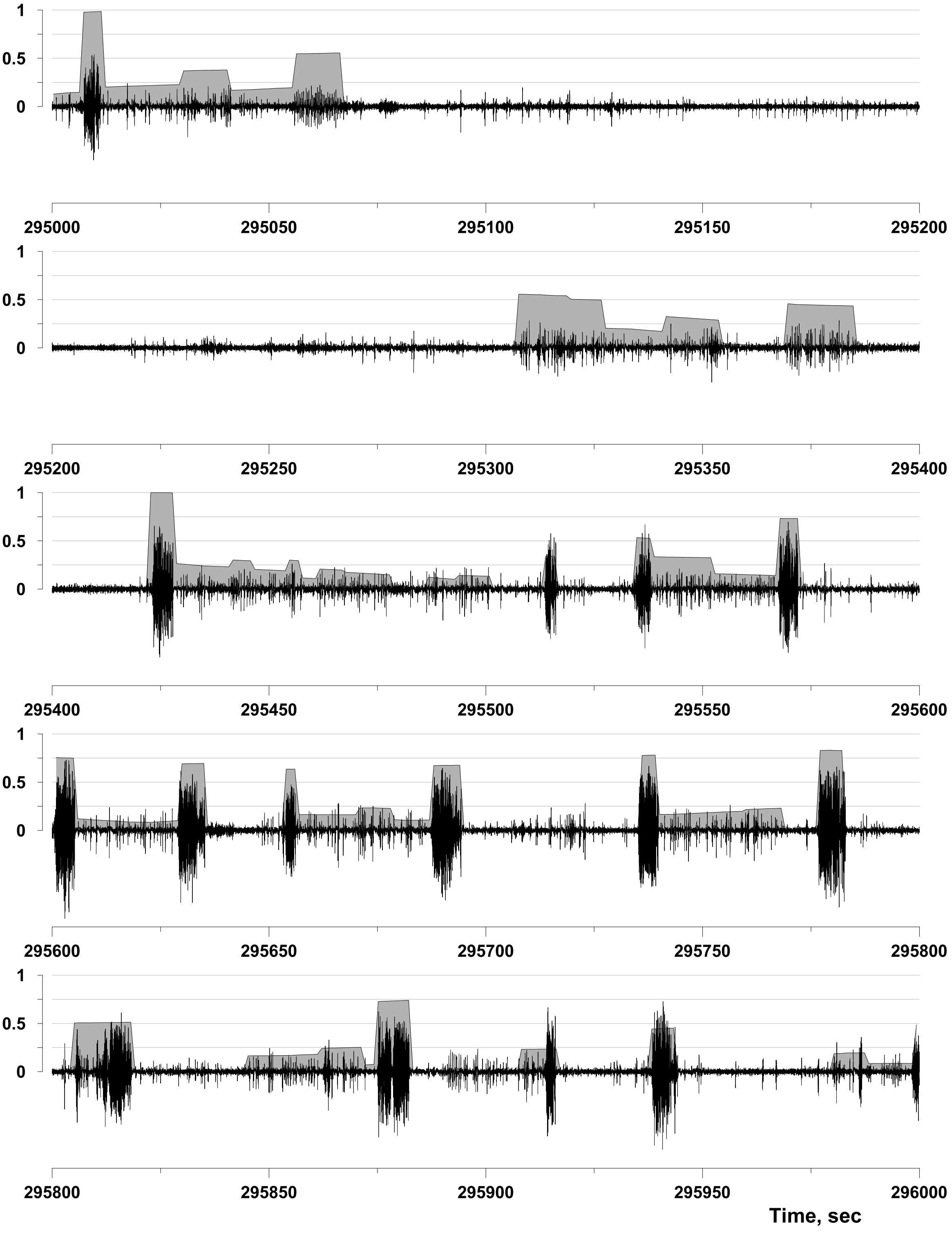

**Figure 3d**

**Figure 3 a-d.** Probability Measure of Seizure Activity estimated using the Wavelet Transform Maximum Modulus - *Stepwise Approximations*. Panels 3a-3d correspond to panels 1a-1d. The oscillations in black are cortical activity and the grey stepwise function, the probability value they correspond to seizures (*x*-axes: time; *y*-axes *PMSA* values.

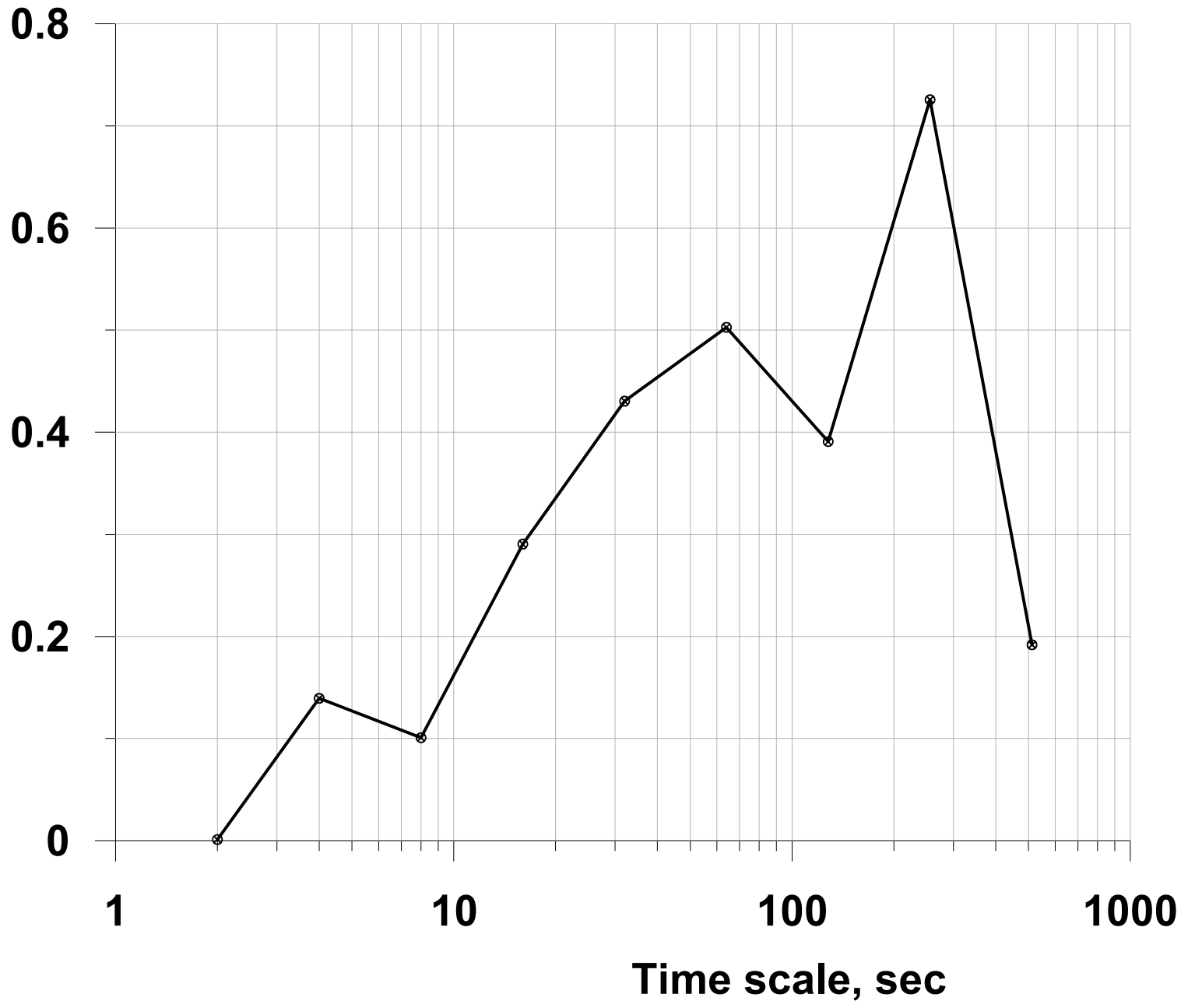

**Figure 4.** Graphic of time scale-dependent correlations between PMSA-*AIF* and PMSA-*SA* after smoothing of their step-wise functions with Haar wavelets. Correlation value increase as a function of time before decaying steeply after approximately 250s.